\documentclass[modern]{aastex631}

\usepackage{amsmath,amsfonts,amssymb}
\usepackage{xcolor}
\usepackage[colorlinks=true]{hyperref}

\shorttitle{Scattering Transform for Roman Dark Images}
\shortauthors{Velicheti, Wu, \& Petric}

\begin{document}

\title{Quantifying Roman WFI Dark Images with the Wavelet Scattering Transform}

\author[0009-0004-2580-3624]{Phani Datta Velicheti}
\affiliation{Department of Mathematics, The University of Arizona, 617 N. Santa Rita Ave., P.O. Box 210089, Tucson, AZ 85721}

\author[0000-0002-5077-881X]{John F. Wu}
\affiliation{Space Telescope Science Institute, 3700 San Martin Dr, Baltimore, MD 21218}
\affiliation{Department of Physics \& Astronomy, Johns Hopkins University, 3400 N Charles St, Baltimore, MD 21218}
\email{jowu@stsci.edu}

\author[0000-0003-4030-3455]{Andreea Petric}
\affiliation{Space Telescope Science Institute, 3700 San Martin Dr, Baltimore, MD 21218}
\affiliation{Department of Physics \& Astronomy, Johns Hopkins University, 3400 N Charles St, Baltimore, MD 21218}

\begin{abstract}
The Nancy Grace Roman Space Telescope will survey a large area of the sky at near-infrared wavelengths with its Wide Field Instrument (WFI).
The performance of the 18 WFI H4RG-10 detectors will need to be well-characterized and regularly monitored in order for Roman to meet its science objectives. 
Weak lensing science goals are particularly sensitive to instrumental distortions and patterns that might masquerade as astronomical signals.  
We apply the wavelet scattering transform in order to analyze localized signals in Roman WFI images that have been taken as part of a dark image test suite.
The scattering transform quantifies shapes and clustering information by reducing images into non-linear combinations of wavelet modes on multiple size scales.
We show that these interpretable scattering statistics can separate rare correlated patterns from typical noise signals, and we discuss the results in context of power spectrum analyses and other computer vision methods.
\end{abstract}
\keywords{Infrared observatories (791), Astronomical detectors (84), Astrostatistics techniques (1886), Observational cosmology (1146)}

\section{Introduction}
\label{sec:intro}

The Nancy Grace Roman Space Telescope (hereafter, Roman) is a NASA flagship mission that will launch in the mid-2020s. 
Roman will be equipped with a Wide Field Instrument (WFI) that will allow it to conduct several core astronomical surveys at visible and near-infrared wavelengths.
The WFI detector system will be used to observe large areas of the sky with high sensitivity and exquisite resolution. 
It features a mosaic of 18 contiguous detectors arranged in a characteristic arc-like pattern in the focal plane, allowing it to map $0.281~{\rm deg}^2$ of sky per pointing.
Each H4RG-10 detector delivers $4096 \times 4096$ pixels at $0.11$~arcsec~pixel$^{-1}$ resolution.

The science goals of the Roman Space Telescope's core surveys heavily depend on the performance and stability of its detectors. 
For example, noise correlations in the data collected by the Roman WFI can mimic weak lensing signals, which could compromise the scientific interpretation of the observational cosmology program.
To mitigate these potential issues, Roman will regularly collect dark exposures and other calibration images during its mission. 
These data will be used to measure the performance of the detectors, allowing the Roman operations team to update reference files or calibration steps as needed.

In order to determine whether the detectors are experiencing \textit{anomalous} behavior, it is necessary to understand the \textit{baseline} detector behavior.
Unfortunately, it is challenging to summarize a 16 megapixel detector in just a few metrics.
Simple statistics such as the mean, median, and standard deviation are able to detect large-scale shifts in the distribution of pixel values, but remain insensitive to rare, aberrant signals that only affect the tail of the distribution.
Moreover, these simple statistics cannot alert us to the presence of spatially correlated structures.
Instead, more advanced methods are needed for extracting statistics on multiple scales and across a range of correlated patterns.

In this work, we analyze Roman ground-test dark images (darks) in order characterize their statistical properties. 
We evaluate a machine learning method called the wavelet scattering transform, which can be used to extract high-level summary statistics from the images.
The wavelet scattering transform summarizes the dark images into various wavelet ``modes'' quantified by their scattering coefficients.
The scattering coefficients provide valuable and interpretable statistics for many astrophysical topics, ranging from observational cosmology \citep[e.g.,][]{Cheng+2020,ChengMenard21a,Greig+22,ValogiannisDvorkin22} to studies of Galactic dust and the interstellar medium \citep[e.g.,][]{Regaldo-SaintBlancard20,Saydjari+21,LeiClark22}.
We find that certain combinations of coefficients, $s_{21}$ vs $s_{22}$, which are first described in \cite{ChengMenard2021}, present useful a useful way to characterize darks.
We compare our results with power spectrum analyses, which are able to extract statistical variance over a range of scales \citep[e.g.,][]{Pullen+16}, but are unable to represent any \textit{shape} information.

Our paper is outlined as follows.
In Section~\ref{sec:data}, we provide an overview of the ground-test data used in the analysis.
In Section~\ref{sec:method}, we describe the wavelet scattering transform, and we present our analysis in Section~\ref{sec:analysis}.
In Section~\ref{sec:results}, we report our results, and finally we discuss implications and future directions in Section~\ref{sec:discussion}.

\section{Dark test imaging data}
\label{sec:data}

Our data comprises dark current images taken as part of the Roman WFI ground tests.
In Section~\ref{sec:dark-patterns}, we describe the WFI dark current science requirements.
In Section~\ref{sec:detectors}, we provide a brief overview of the properties of the WFI detectors.
In Section~\ref{sec:darks}, we describe the measurement of dark current rates, which are used to produce the dark images in our analysis.

\subsection{Sources of noise and correlated patterns} \label{sec:dark-patterns}

Dark signal can arise from internal detector currents, residual biases, and thermal background. Weak lensing studies with Roman require that the dark signal be known to 0.0114 $e^{-}$\,s$^{-1}$\,pixel$^{-1}$ in the absence of noise correlations \citep{WFIRST}. For a 300\,s dark exposure in imaging mode, the required $1\,\sigma$ knowledge of dark signal is $0.007$~$e^{-}$\,s$^{-1}$\,pixel$^{-1}$. Such stringent calibration needs suggest that the stability of dark current must be regularly monitored and characterized. Spatial structures in the dark images may indicate noise correlations that could decrease the allowable uncertainty on pixel-level dark current knowledge. In this work, we propose to use a wavelet scattering transform to identify these correlated patterns in Roman WFI darks.

\subsection{WFI H4RG-10 Detectors}\label{sec:detectors}

The Roman H4RG-10 detectors are mercury-cadmium-telluride (HgCdTe) arrays produced by Teledyne.
They are similar in design to the earlier-generation infrared detectors, which reside aboard the Hubble Space Telescope (WFC3) and JWST (NIRCam, NIRSpec, and FGS). 
The JWST mission is presently producing real calibration and science data at L2, enabling better characterization of H2RG performance in the true environment \citep[e.g.,][]{2023PASP..135d8001R}.

We use 44 darks taken by the Goddard Space Flight Center's Detector Characterization Lab as part of a set of tests to investigate count rate non-linearity \citep{Mosby+2020} in one of the 18 detectors that will make up Roman's focal plane system. The data files used in the analysis are darks taken with detectors connected to non-flight electronics (e.g., using the Leach controllers rather than the ACADIA controllers). 

\cite{Mosby+2020} provide a comprehensive overview of the H4RG-10 detectors (also known as sensor chip assemblies, or SCAs) and their expected properties when cooled and stabilized at a temperature of 100~K. 
The Roman detectors have high quantum efficiency and low read noise, persistence, non-linearity (classical and count rate), inter-pixel capacitance, burn-in, and crosstalk. 
Here we focus on the contribution of dark current structures on potential science measurements, and direct our investigation toward the two-dimensional structures in the darks. 

\subsection{Measuring Dark Current Rates} \label{sec:darks}

The Roman IR detectors use non-destructive readouts, such that the accumulated charge can be repeatedly read out during an exposure. The amount of charge versus time is sometimes referred to as a ramp in WFC3, NIRCam, and Roman literature. The rate of accumulating signal in each pixel can be computed by estimating the slope of the ramp, a method known as up-the-ramp fitting \citep[e.g.,][]{Fixsen+2000,Casertano2022}.

Each dark file used for this analysis is organized in the form of a ramp made up of a number $N_r$ of reads separated by $N_{\rm skip}$ reads, each frame/read being read after a time $t_{\rm fr}$.
The integration time $t$ associated with each read $r$ is then given by $$ t = t_{\rm fr}+(N_{\rm skip}+1) \cdot r \cdot t_{\rm fr} .$$
For each read, we perform a reference pixel correction and subtract the first read. Next, for each pixel location, we fit a line to the accumulated dark signal at each read as a function of the integration time. We do not include the first two reads in the fit to avoid bias fluctuations that are sometimes present at the start of an observation. 
We assign a weight to each pixel in each read based on the read-noise estimates for that particular pixel. We estimate the read noise from the correlated double sampling (CDS) noise: if we define $F_r(\vec x)$ to be the signal in pixel position $\vec x$ and read number $r$, then we can compute an array of uncorrelated differences, i.e., $[F_r(\vec x)-F_{r-1}(\vec x)] - [F_{r+2}(\vec x) - F_{r+1}(\vec x)]$. For each pixel, we estimate the standard deviation of all the differences computed from pairs of reads, and we compute the read noise as this standard deviation divided by $\sqrt{2}$ \citep[][]{Petric+23}.

\section{The wavelet scattering transform}
\label{sec:method}

The wavelet scattering transform is a method for summarizing the statistical properties of an input signal, such as an image or a field \citep{Mallat2012}.
It is analogous to the power spectrum $P(\vec k)$, which describes the variance of a signal as a function of spatial frequency $\vec k$.
However, the power spectrum is limited to probing only the variance of fluctuations at different scales, as the power spectrum eschews the input signal's Fourier phases in favor of the squared amplitudes.
Meanwhile, the wavelet scattering transform is designed to extract complex \textit{morphological} information from the phases, producing a richer characterization of the input signal's power and \textit{shape} at multiple size scales and shape orientations.
The properties are critical for understanding why convolutional neural networks (CNNs) are so effective at extracting information from input signals \citep{BrunaMallat2013,Mallat2016}.

Given an input signal $I(\vec x)$, the wavelet scattering transform performs convolution and modulus operations, in that sequence: $I(\vec x) \rightarrow |I(\vec x) * \psi(\vec x) |$.
In our work, $\psi$ is a Morlet wavelet, e.g. a two-dimensional sinusoidal pattern in a Gaussian envelope.
We also note that $\vec x$ here is defined over the discrete grid of pixels, although the scattering transform formalism can also be extended to the continuous limit \citep{Mallat2012}.
The $n$th order scattering coefficients $S_n$ serve as summary statistics, and they are defined as the spatial average of the transformed signal (denoted as angular brackets below).
Here we define the zeroth, first, and second order scattering coefficients:
\begin{align}
    S_0 =&~  \langle I(\vec x) \rangle, \\
    S_1^{(j_1, l_1)} =&~  \big\langle  \big\vert I(\vec x) * \psi^{(j_1,l_1)}(\vec x) \big\vert \big\rangle, \\
    S_2^{(j_1, l_1, j_2, l_2)} =&~  \big\langle \big\vert \vert I(\vec x) * \psi^{(j_1,l_1)}(\vec x) \vert * \psi^{(j_2,l_2)}(\vec x)\big \vert \big \rangle, 
\end{align}
where $\psi_{j,l}$ is a family of wavelets parameterized by a (logarithmic) size $j$ and orientation $l$.
It is useful to perform the scattering transform over a range of scales $2^j \in \{ 2^1, 2^2, \ldots, 2^J\}$.
Similarly, a range of orientation angles $l \times \pi/ L$ can be defined for $l \in \{1, 2, \ldots L\}$.
We select the same $J$ and $L$ hyperparameters for first- and second-order coefficients.

In summary, the scattering transform matches wavelet patterns across multiple sizes and orientations through a series of convolutions: $I(\vec x) * \psi^{(j,l)}(\vec x) $. 
The modulus serves as a non-linear operator that ``scatters'' information from higher-frequency modes to lower-frequency modes, producing coefficients that are \textit{locally} invariant to translation \citep{Guth+22}.
The spatial average allows us to extract the first-order statistics, or a second round of the scattering transform can be performed to identify yet higher-frequency information. 
The second-order coefficients quantify the clustering strength of $j_1$-sized features on scales defined by $j_2$. 
Due to symmetry, it is only useful to consider $j_2 > j_1$.
We refer the interested reader to \cite{ChengMenard2021}, who present a guide for understanding the physical intuition behind the wavelet scattering transform.

Because the number of coefficients rapidly increases with order, i.e. one zeroth-order coefficient, $J \times L$ first-order coefficients, and $J(J-1)/2 \times L^2$ unique second-order coefficients, it is useful to summarize this information.
One option would be to use a dimensionality reduction technique, e.g., Principal Components Analysis (PCA), Independent Components Analysis, (ICA), $t$-distributed Stochastic Neighbor Embedding (t-SNE), or Uniform Manifold Approximation and Projection (UMAP).
Alternatively, it is possible to define meaningful summary statistics from the scattering coefficients by taking selective averages.
We adopt the $s^{(j_1,j_2)}_{21}$ and $s^{(j_1,j_2)}_{22}$ statistics introduced by \cite{ChengMenard2021}:
\begin{equation}
    s^{(j_1, j_2)}_{21} = \left \langle S^{(j_1, l_1, j_2, l_2)}_2 / S^{(j_1,l_1)}_1 \right \rangle_{l_1, l_2},
\end{equation}
which provides a notion of \textit{sparsity}, and 
\begin{equation}
    s^{(j_1, j_2)}_{22} = \left \langle S^{(j_1,j_2,l_1 = l_2)}_2 / S^{(j_1,j_2,l_1 \perp l_2)}_2 \right \rangle_{l_1},
\end{equation}
which can be thought of as a shape index, or ``filamentarity.'' 
Here the $l_1 = l_2$ and $l_1 \perp l_2$ expressions select parallel and perpendicular orientations from the possible values of $l$, respectively.\footnote{An even value of $L$ is required to compute the $s^{(j_1, j_2)}_{22}$ statistics.}
After taking the averages over orientation angles, there remain $J(J-1)/2$ unique $s^{(j_1,j_2)}_{21}$ statistics and another set of $J(J-1)/2$ unique $s^{(j_1,j_2)}_{22}$ statistics.
Since we are not \textit{a priori} searching for correlated signals at specific scales, we average over all $j_1$ and $j_2$, which allow us to further compress the scattering statistics, and hereafter we will write the averaged statistics without superscript as $s_{21}$ and $s_{22}$.

\begin{figure}
    \centering
    \includegraphics[width=\columnwidth]{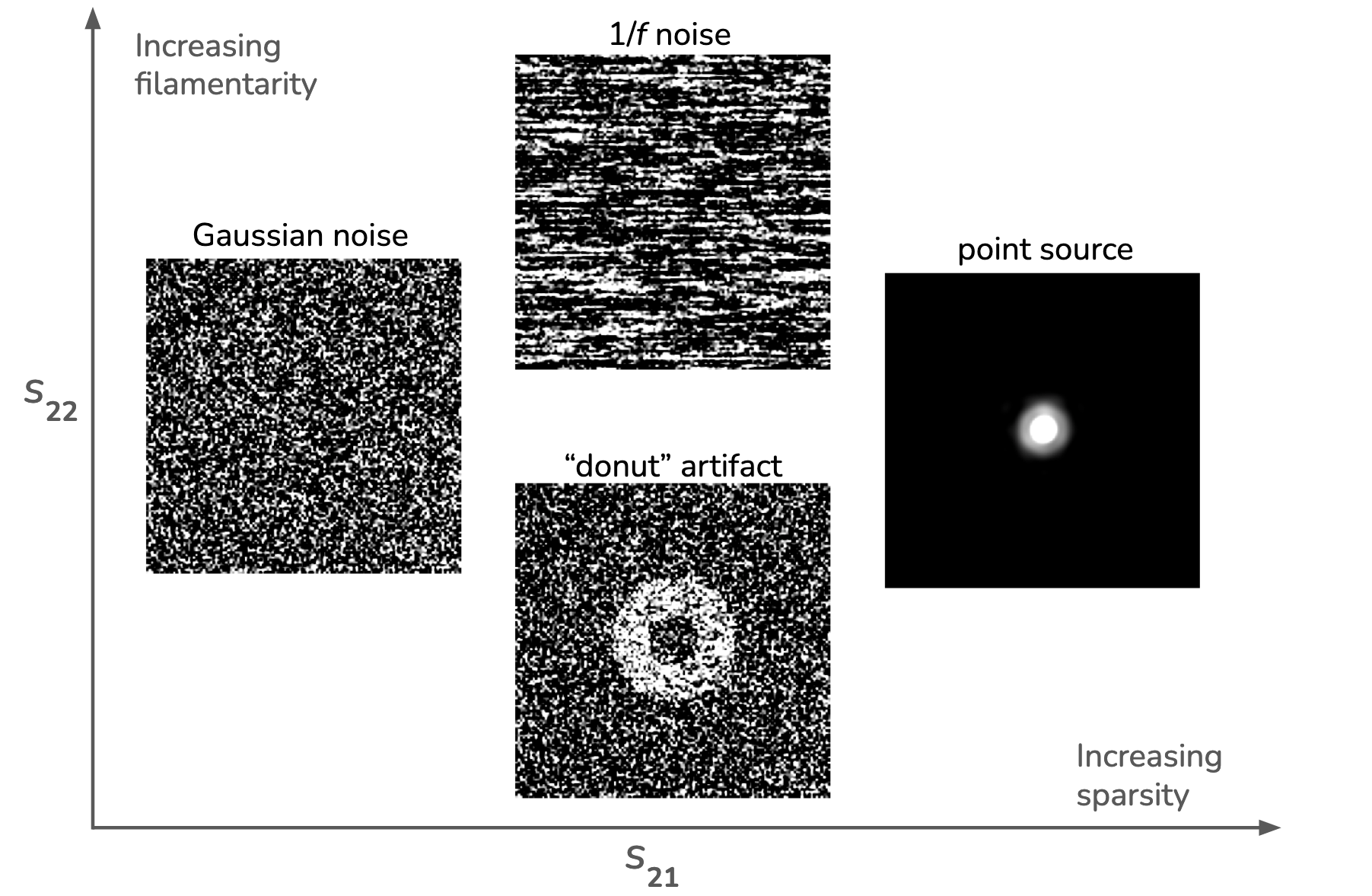}
    \caption{Example $128 \times 128$ images of Gaussian noise, $1/f$ noise, a ``donut'' artifact on Gaussian noise, and a point source function. 
    Each image is placed in a different qualitative region of $s_{21}$--$s_{22}$ space.}
    \label{fig:toy-examples}
\end{figure}

In order to provide some intuition for the $s_{21}$ and $s_{22}$ statistics, we illustrate several toy examples in Figure~\ref{fig:toy-examples}.
We simulate image cutouts displaying (i) isotropic Gaussian-distribute noise, (ii) 1/$f$ noise along the detector rows, (iii) a ``donut'' shaped artifact on top of isotropic Gaussian noise,\footnote{We note that the donut patterns seen in the data tend to be much smaller (see, e.g., Section~\ref{sec:results} and Figure~\ref{fig:representative-examples} than the example we show here; our primary objective for simplified illustrations is to build intuition rather than simulate realism.} and (iv) a single point source in the F146 filter generated from a effective point spread function (ePSF) model \citep[using WebbPSF;][]{WebbPSF}.
We find that the pure Gaussian noise resides in the expected locus of $s_{21}-s_{22}$; the noise patterns exhibit no real clustering, resulting in low $s_{21}$, and no morphological information exists (since a Gaussian random field can be exactly described by a power spectrum), resulting in neither filament-like nor curve-like structures (intermediate $s_{22}$).
The $1/f$ noise manifests as horizontal correlated patterns along the fast-read direction, and therefore results in a higher $s_{22}$ statistic due to its filamentary appearance.
The donut artifact is a curved pattern, and conversely results in a lower $s_{22}$ statistic.
Both of these structures are more sparse than a Gaussian random field and correspondingly have higher $s_{21}$.
Finally, the simulated point source is very sparse, resulting in high $s_{21}$, and it has very little morphological information and therefore intermediate $s_{22}$.
Using these toy examples as intuitive guides, we interpret $s_{21}$ the feature sparsity parameter and $s_{22}$ the filamentarity parameter.

\section{Applying the scattering transform to Roman darks} \label{sec:analysis}

We are now ready to characterize the statistical properties of the Roman dark images.
In order to compute the scattering transform, we must select the $J$ and $L$ hyperparameters.
We choose to run our analysis using a maximum scale defined by $J=7$.
This cutoff helps reduce the computational cost of evaluating convolutions using large wavelet kernels.
We are also primarily interested in finding correlated signals on small scales, given that very large-scale spatial correlations can be visually identified rather easily, whereas small-scale correlations are harder to find.
In a later section (Section~\ref{sec:large-scales}), we compare the impact of restricting our analysis to smaller scales ($J = 7$) versus extending it to much larger scales (analogous to $J = 12$).
We also evaluate the impact of selecting $L= 4$, $6$, and $8$.
Based on a comparison of the $s_{21}$ and $s_{22}$ statistics, we find essentially no difference in the overall trends as we vary $L$.
Indeed, the scattering coefficients will be correlated with each other beyond $L>2$, so this finding is no surprise.
Therefore we proceed with our analysis using $J=7$ and $L=4$.

Because we are only considering structures on scales smaller than $2^J = 128$ pixels, we split each dark image into 1024 patches of size $128 \times 128$ pixels.
We note that these $128 \times 128$-pixel patches are also sometimes called ``superpixels'' in the literature \citep[e.g.,][]{Troxel+23}.
This preprocessing step is also appropriate since gain corrections will be applied for each amplifier, which reads out vertical columns of $128 \times 4096$ pixels, and thereby inducing correlations on strips of width $128$ pixels. 
Additionally, this step enables us to process many image patches in parallel.
For the details of the reference pixel corrections used to derive the dark darks for this analysis, see \cite{Petric+23}.
We also note that the patches along the border of each dark image have reference pixels, which are used to measure and remove correlated noise, such as $1/f$ noise, over the entire detector image \citep[e.g.,][]{Rauscher+17}.
A pedestal can sometimes appear between the reference pixels and the rest of the detector pixels, which manifests as a sharp morphological feature (i.e. an edge or filament-like feature); we find that this preferentially pushes all border pixels to high $s_{22}$ values.
For this reason, we remove all dark image border patches from our analysis, resulting in $900$ patches per image.
In summary, we compute local scattering statistics, $s_{21}, s_{22}$, for each dark image patch that is not located on the image border.

We compute the wavelet scattering transform using the \texttt{scattering} package\footnote{\url{https://github.com/SihaoCheng/scattering_transform}} (which is based on the \texttt{kymatio} package; \citealt{kymatio}). 
The scattering transform employs the Morlet family of wavelets (see Appendix B of \citealt{ChengMenard2021} for a discussion on Morlet wavelets).
The $s^{(j_1,j_2)}_{21}$ and $s^{(j_1,j_2)}_{22}$ statistics can also be generated using the \texttt{scattering} package.
Since we are generally concerned with the qualitative trends of the averaged $s_{21}$, $s_{22}$ statistics, we report only their relative values (i.e. compared to the mean statistic).

\section{Results} \label{sec:results}

\subsection{Quantifying image patch morphologies} \label{sec:scattering-results}

\begin{figure}
    \centering
    \includegraphics[width=\textwidth]{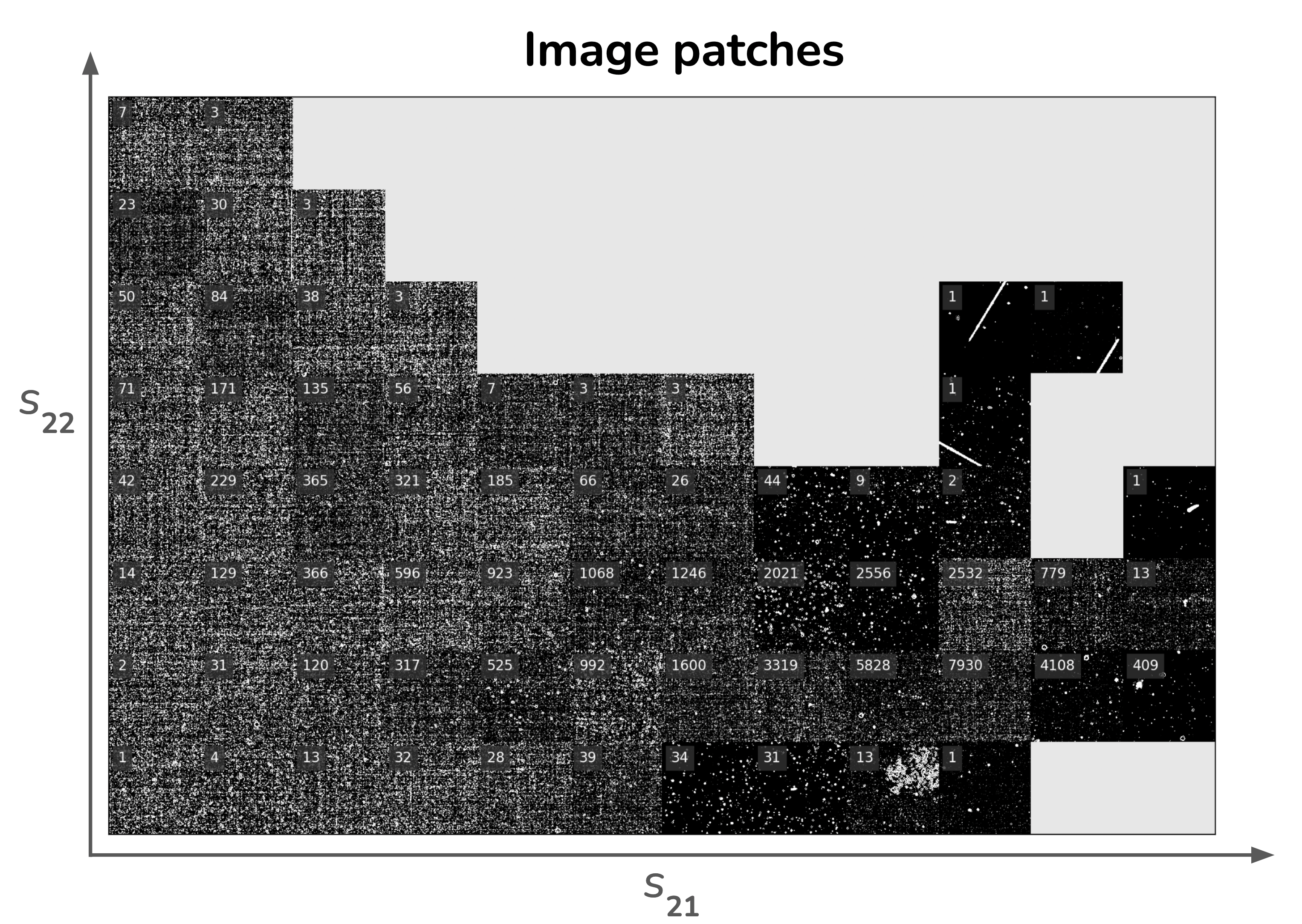}
    \caption{Randomly sampled image patches from bins in the $s_{21}$--$s_{22}$ plane. Gray regions show combinations of $s_{21}$--$s_{22}$ that are not present in our darks data set. The numbers denote how many image patches are in each $s_{21}$--$s_{22}$ bin. 
    }
    \label{fig:s21-s22-images}
\end{figure}

\begin{figure}
    \centering
    \includegraphics[width=\textwidth]{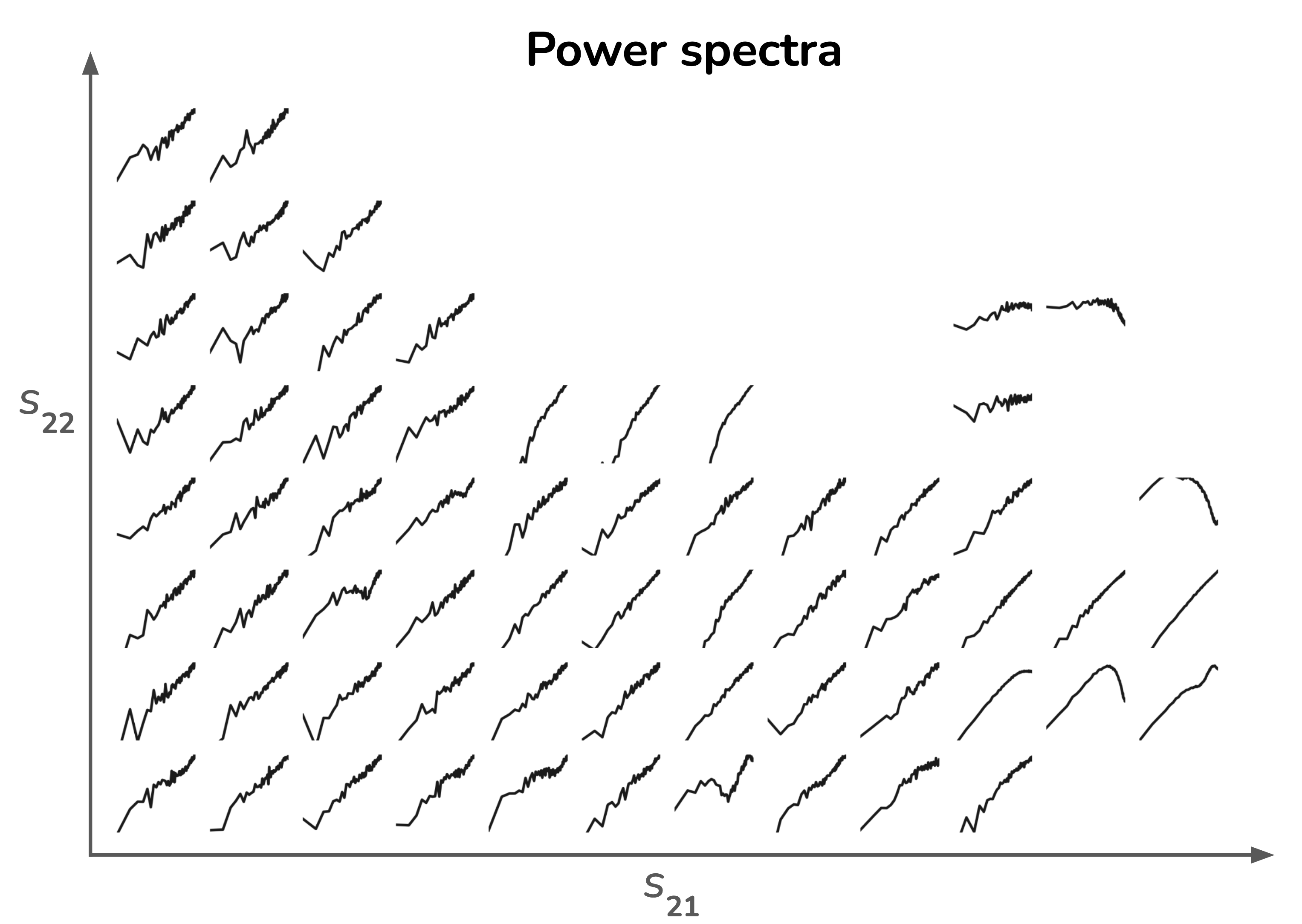}
    \caption{Power spectra for each of the corresponding image patches shown in Figure~\ref{fig:s21-s22-images}. 
    All power spectra have the same axes as in Figure~\ref{fig:representative-examples}. 
    }
    \
    \label{fig:s21-s22-powerspectra}
\end{figure}

We compute scattering statistics for each of the $44 \times 900=$~39,600 patches that are not along the darks image borders. 
In Figure~\ref{fig:s21-s22-images}, we display image patches corresponding to $s_{21}$--$s_{22}$ values, where we have subtracted the sample-averaged statistics, $\langle s_{21}\rangle$ and $\langle s_{22}\rangle$.
We bin the image patches by their $s_{21}$--$s_{22}$ values, and randomly select a patch representing each locus.
The image patches are all shown using the same grayscale linear stretch.
Some regions are shown in light gray because those $s_{21}$--$s_{22}$ statistics do not describe any of our darks.
The figure also displays the number of patches that belong to each $s_{21}$--$s_{22}$ bin.
Image patches with high $s_{21}$ but relatively low values of $s_{22}$ are most common, and do not seem to exhibit many interesting features.

We find that different morphological features appear in different regions of the $s_{21}$--$s_{22}$ plane.
For example, images patches with lower $s_{21}$ tend to exhibit the cross-hatching pattern that is common in the darks \citep[likely due to $1/f$ and read noise; we note that these correlated noise patterns can be mitigated, e.g.,][]{Rauscher+12}. 
Patches with higher $s_{21}$ and relatively high $s_{22}$ show filament-like features, reminiscent of long cosmic ray trails.
Patches with higher $s_{21}$ and lowest $s_{22}$ show bubble- or donut-like features.
In order to distinguish the filament-like and bubble-like artifacts, we need to employ cuts along both $s_{21}$ and $s_{22}$.

\subsection{Comparison to power spectra} \label{sec:power-spectra}

In Figure~\ref{fig:s21-s22-powerspectra}, we show a representative power spectrum for each corresponding image patch shown in Figure~\ref{fig:s21-s22-images}.
The power spectra are computed by taking the squared amplitude of the Discrete Fourier Transform for each image patch.
The Discrete Fourier Transform operates under the assumption of periodic boundary conditions; image patches do not truly have periodic boundary conditions, but this simplification is approximately true if the patch is much larger than the image features in question.
For ease of comparison, the power spectra in Figure~\ref{fig:s21-s22-powerspectra} are displayed in log-log space using consistent $x$- and $y$-axis scales.

We find that there are similarities between power spectra and $s_{21}-s_{22}$ statistics for the images shown in Figure~\ref{fig:s21-s22-images}.
The $s_{21}$ statistic can be interpreted as a measure of sparsity, so sparse (and rare) patterns tend to shown up on the right-hand side of the figure.
Figure~\ref{fig:s21-s22-powerspectra} demonstrates that many high-$s_{21}$ image patches also result in anomalous power spectra.

\begin{figure}
    \centering
    \includegraphics[width=0.495\textwidth]{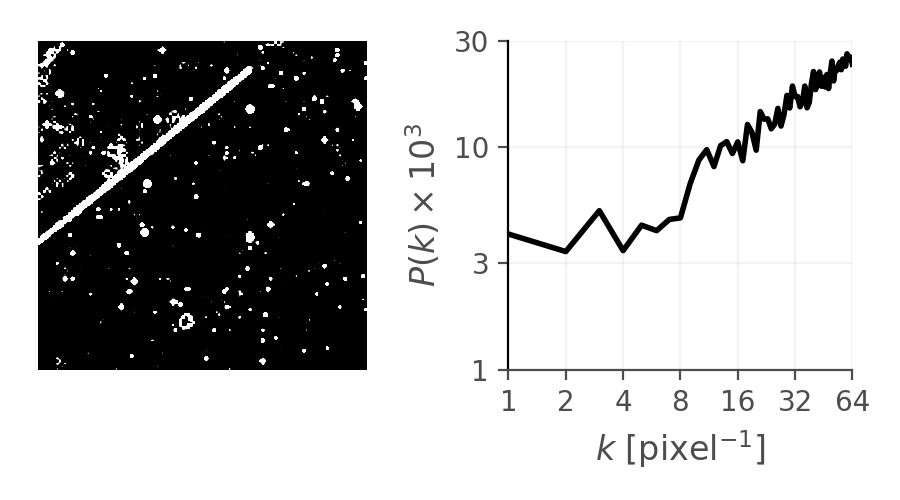}
    \includegraphics[width=0.495\textwidth]{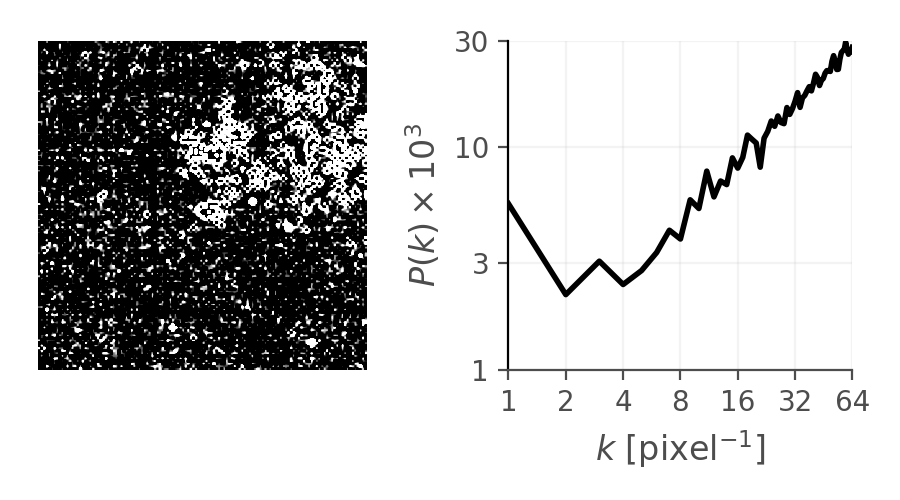}
    \includegraphics[width=0.495\textwidth]{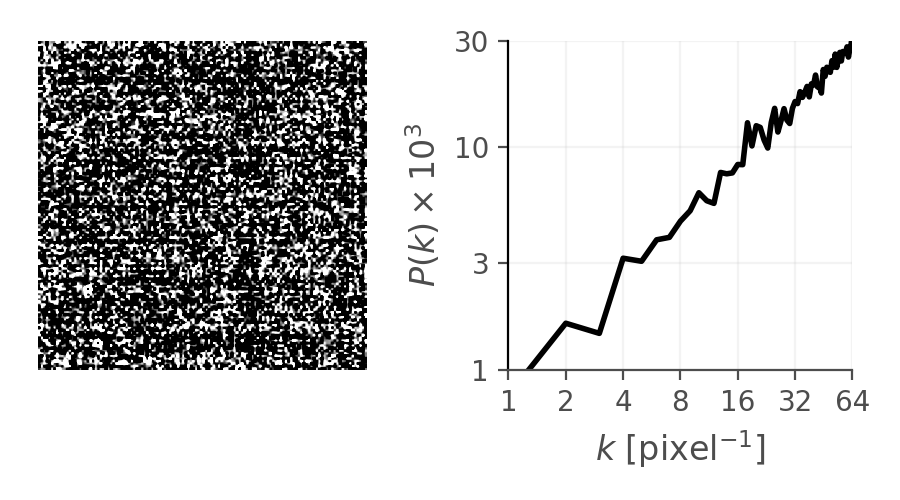}
    \includegraphics[width=0.495\textwidth]{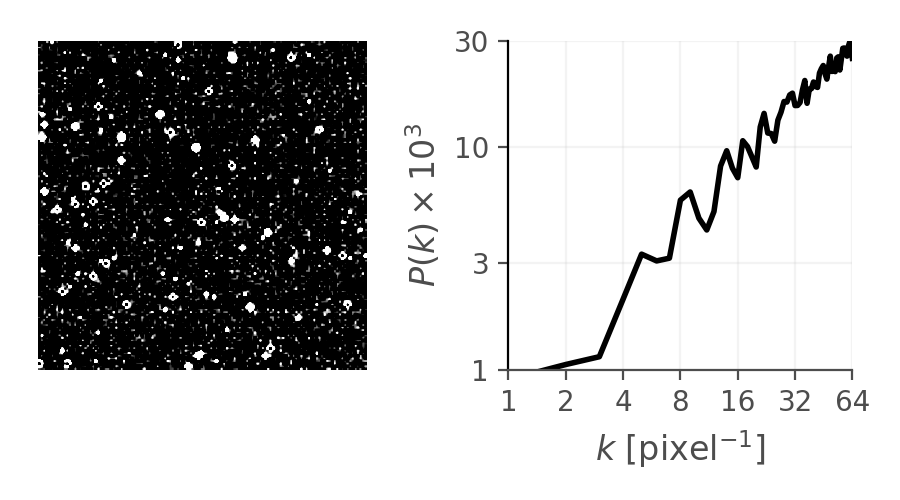}
    \caption{
    $128 \times 128$-pixel darks image patches with corresponding power spectra.
    Each set depicts a representative example of a particular locus in $s_{21}$ (sparsity) versus $s_{22}$ (filamentarity) space.
    \textit{Upper left}: High sparsity and high filamentarity example, showing a rare linear-feature artifact.
    \textit{Upper right}: Medium sparsity, low filamentarity example, showing a rare cluster of donut artifacts.
    \textit{Lower left}: Low sparsity, intermediate filamentarity example, showing common noise and cross-hatched patterns.
    \textit{Lower right}: High sparsity, low filamentarity example, showing common donut artifacts. 
    }
    \label{fig:representative-examples}
\end{figure}

In Figure~\ref{fig:representative-examples}, we show representative examples of image patches and power spectra from different regions in $s_{21}-s_{22}$ space.
Four types of correlated signals are shown here: a single bright filamentary structure (\textit{upper left)}, a tight cluster of small donut artifacts (\textit{upper right}), a typical noise pattern (\textit{lower left}), and an unclustered patch of donut artifacts (\textit{lower right}).
Although these image patches are visually distinct, and are well-separated in the $s_{21}-s_{22}$ space, they do not have particularly unique power spectra.
These examples highlight the benefit of using scattering statistics: different kinds of morphological patterns are intuitively separated in an interpretable manner.

\subsection{Large-scale features}  \label{sec:large-scales}

Our method allows us to probe correlated signals at scales between 1 and 128 pixels. 
We have purposely restricted the range of scales in order to easily visualize the image patches.
However, larger-scale correlations will be missed by our approach.
Therefore, we also repeat our analysis over larger scales: we take the average over bins of $32 \times 32$ pixels, resulting in a single $128 \times 128$ dark image, and then apply the wavelet scattering transform over this downsampled image.
This is analogous to applying a low-pass filter, or repeating our analysis with only $j=\{5, 6, 7, \ldots, 12\}$; however, these operations are not precisely equivalent, because the scattering transform captures interactions across two different sets of scales and orientations.

We find that the scattering transform on these downsampled images is not as informative: interesting small-scale signals vanish, while large correlated patterns are already discernable through visual inspection.\footnote{A two-dimensional Fourier-Transform power spectrum analysis of spatial structures also shows systematic differences between detectors \citep{Petric+23}.} 
Moreover, all 44 downsampled dark images can be quickly viewed in a short amount of time, bypassing the need for any statistical or machine learning tools.
The scattering transform and power spectrum still offer tools for quantifying these large-scale features, which may be useful in other contexts, but we find that such analyses are more valuable for finding small features.
Therefore, we continue our scattering transform analysis to only the smaller image patches.

\section{Discussion} \label{sec:discussion}

In this paper, we have presented a method for characterizing the ground-test dark data using the wavelet scattering transform. 
The mathematical formulation of our methodology is described in Section~\ref{sec:method}.
For our analysis (Section~\ref{sec:analysis}), we split dark images into $128 \times 128$-pixel patches and measured their summarized wavelet scattering statistics, $s_{21}$ and $s_{22}$.
These scattering statistics quantify distinct types of correlated signals, such as feature sparsity or linearity (see Section~\ref{sec:results} for our results).

Power spectra can also characterize the signal strength at various size scales \citep[e.g.,][]{Petric+23}, but they are limited in picking up morphological features contained in the Fourier phases (see Section~\ref{sec:power-spectra}; for more discussion, see also  \citealt{2019ApJ...882L..12P}). 
Circular or linear features can also be identified via other methods such as Hough Transforms \citep[e.g.,][]{2006AdSpR..37...21S,2014ApJ...789...82C}.
However, there are several drawbacks to using these algorithms; for example, specialized Hough Transforms are necessary for detecting different shapes.
Therefore, a library of detection algorithms need to be tuned based on a subset of the data---i.e. a training set---but these manually selected algorithms may not be sensitive to features in previously unseen data (i.e., a test set).
Alternatively, deep convolutional neural networks operate in a similar manner to the wavelet scattering transform \citep[e.g.,][]{BrunaMallat2013}, and pre-trained networks may be used to separate image signals in an unsupervised fashion \citep{2020ASPC..522..381P}.
Unfortunately, convolutional neural networks tend to decompose images into a large number ($\sim 1000$) of uninterpretable morphological features.
Meanwhile, the wavelet scattering transform is a sensitive to image morphology, can be used without a training set, and reduces images into interpretable statistics.

We emphasize that the $s_{21}$--$s_{22}$ statistics are practical for separating common noise patterns from rare and potentially impactful signals.
Our analysis covered only 44 dark images, split up into 39,600 non-border image patches, and all tests have been performed on a single Roman WFI detector.
While undertaking this analysis, the authors visually inspected the images for interesting examples of correlated signals and detector artifacts.
However, only $\sim 0.2$\% of the image patches exhibit filament-like and bubble-like features, and thus our painstaking visual inspection campaign was very inefficient and resulted in few useful examples.
After computing the scattering statistics and visualized the results (Figure~\ref{fig:s21-s22-images}), we were immediately able to spot these features and dramatically narrow down the search space.
In the future, far larger suites of WFI test data or on-flight reference data will be taken as part of the Roman mission, and these scattering statistics will be invaluable for efficiently identifying anomalous signals.

Other than dark current, there are many instrumental effects or physical processes that can obfuscate the measurement of faint astronomical signals. 
For example, the Roman WFI will need to account for read noise, $1/f$ noise, persistence, saturation, hot pixels, classical non-linearity, geometric distortions, inter-pixel capacitance, charge diffusion, the brighter-fatter effect, and jumps due to cosmic ray events or snowballs \citep[e.g.,][]{2020PASP..132a4502C,2020PASP..132g4504F,Mosby+2020,2022PASP..134a4001G,2022PASP..134k5001H,2023PASP..135d8001R}.
Future explorations are still needed to determine whether there are other dominant types of detector noise or problematic signals.
While our work focuses solely on darks, we envision that the wavelet scattering transform may also be useful for characterizing other types of signals, particularly spatially or temporally correlated patterns.

\software{
    Astropy \citep{astropy},
    matplotlib \citep{matplotlib},
    Numpy \citep{numpy},
    pandas \citep{pandas},
    Pytorch \citep{pytorch},
    scattering \citep{ChengMenard2021},
    SciPy \citep{scipy}
}

\vspace{1em}
\begin{acknowledgments}
The authors thank the anonymous referee for providing helpful comments.
The authors thank Carolyn Slivinski for organizing the 2022 virtual Space Astronomy Summer Program (SASP) at STScI, where this project first began.
The authors also thank Russell Ryan, whose early contributions helped define the project, and Andrea Bellini, who provided a supersampled version of the Roman WFI effective PSF.
\end{acknowledgments}

\bibliography{main} 
\bibliographystyle{aasjournal}
\end{document}